\documentclass[twocolumn,notitlepage,tightenlines,preprintnumbers,amsmath,amssymb]{revtex4}

\usepackage{hyperref}
\usepackage{graphicx,amsmath}
\usepackage{epsf}
\usepackage{xcolor}

\begin{document}
\title{Squeezed States of Coupled Photons and Phonons in Nanoscale Waveguides}
\author{Hashem~Zoubi}
\email{hashemz@hit.ac.il}
\affiliation{Department of Physics, Holon Institute of Technology, Holon 5810201, Israel}
\date{12 December, 2018}

\begin{abstract}
We investigate the generation of photon-phonon entangled states by exploiting stimulated Brillouin scattering in nanoscale waveguides. The squeezing type Hamiltonian that represents creation and annihilation of photon-phonon pairs is obtained out of the multimode photon-phonon interaction Hamiltonian in the presence of a pump field. The Bogoliubov eigenstates of the Hamiltonian, which are a coherent mix of propagating photons and phonons, show squeezing properties in the uncertainties of the collective quadrature operators. The measurement of the photonic component can provide a controllable source of single phonons in demand.
\end{abstract}


\maketitle

\section{Introduction}

The creation of entanglement between optical and mechanical components is of big importance for quantum information processing in optomechanical setups \cite{Aspelmeyer2014,Clerk2010}. Nanoscale optomechanical systems are a kind of hybrid systems that combine optical and mechanical parts within the same setup \cite{Safavi2019}. The production of various quantum states of the mechanical and optical modes can be achieved with the aid of photon-phonon interactions within nanoscale structures \cite{Mancini1994,Hofer2011}. Interfacing single photons with single phonons allows the quantum states of mechanical motion to be prepared and manipulated by counting the light quanta. The generation of single phonon can be of importance for quantum information and metrology with applications in quantum communications and quantum memories \cite{Riedinger2016,Safavi2019}. The formation of squeezed states in optomechanical systems is a difficult mission due to the influence of dissipation and decoherence \cite{Mari2009,Liao2011,Farace2012}. The unwanted thermal fluctuation of the mechanical modes can be removed using optomechanical cooling senarios \cite{Szorkovszky2011,Schmidt2012}. Moreover, one can appeal to measurement techniques in order to manipulate quantum states of sound in nanostructures \cite{Paternostro2007,Vitali2007,Palomaki2013}.

The generation of squeezed light with fluctuations below the vacuum ones has been of great interest since its theoretical prediction \cite{Yuen1976}. Squeezed states of light has been demonstrated experimentally in different systems ranging from bulk optical nonlinearity down to micromechanical resonators \cite{Safavi2013}. Quantum states of mechanical motion with a well-defined phonon number requires fast quantum operations and long coherence times. Squeezed state of motion in micromechanical resonator has been manipulated using microwave frequency radiation pressure \cite{Wollman2015}. The controlled generation of multi-phonon Fock states in a macroscale bulk acoustic-wave resonator has been demonstrated in \cite{Chu2018}. Non-classical correlations between photons and phonons in a nanomechanical resonator have been reported according to a full quantum protocol involving initialization of the resonator in its ground state of motion and subsequent generation and read-out of correlated photon-phonon pairs \cite{Riedinger2016}, following a probabilistic scheme based on Duan, Lukin, Cirac, and Zoller protocol \cite{Duan2001}. Moreover, the entanglement generation in cavity quantum optomechanics has been implemented in different hybrid quantum systems \cite{Stannigel2010,Pirkkalainen2015,Lemonde2016,Mirza2016,Riedinger2018}.

Stimulated Brillouin Scattering (SBS) is the scattering of light from mechanical excitations of a medium \cite{Boyd2008,Agrawal2013}. The recent progress in fabricating waveguides with nanoscale cross section opens new horizons for
SBS. Here a breakthrough has been appeared in which radiation pressure dominates over the conventional electrostriction mechanism, as predicted theoretically \cite{Rakish2012} and realized later
experimentally \cite{VanLaer2015a}. Radiation pressure can provide strong photon-phonon interactions that
lead to a significant enhancement of SBS in nanoscale waveguides \cite{Zoubi2016}. Such a progress introduces SBS as a promising candidate for quantum information processing involving photons and phonons in nanoscale structures \cite{Thevenaz2008,Bahl2012,Agarwal2013a,Beugnot2014,Merklein2016,Zoubi2017,Zoubi2018}.
Various setups have been fabricated for the realization of nanoscale
waveguides \cite{Eggleton2013}. The low waveguide
mechanical quality factor is among the main
factors that strongly limit the efficiency of each device \cite{Safavi-Naeini2011,Weis2010,Kim2015,Aspelmeyer2014}. For example, in on-chip waveguides the direct contact between the waveguide and the substrate
material limits the quality factor to small values with a relatively fast leak of the phonons into the substrate \cite{Pant2011}. On the
other hand, suspended nanoscale
waveguides have higher quality factor with longer phonon
lifetime, but they are limited to a very short waveguide length that yields weak SBS, e.g. in silicon nanowires \cite{Shin2013}. A compromise has been suggested by using nanoscale
silicon photonic wires that are supported with a tiny pillar, the fact that keeps a reasonable
quality factor and allows the achievement of relatively long waveguides with strong SBS \cite{VanLaer2015a,VanLaer2015b,Kittlaus2015}.

In the present paper we use our previous results on continuum quantum optomechanics \cite{Zoubi2016} in order to generate entangled photons and phonons with squeezing properties in nanstructures. The process of the creation and annihilation of photon-phonon pairs can be achieved in exploiting SBS in nanoscale waveguides. Such propagating photons and phonons obey conservation of energy and momentum in the presence of a strong pump field. The quadratic Hamiltonian obtained in assuming a classical pump field can be diagonalized using the known Bogoliubov transformation. The diagonal eigenstates show strong entanglement between photons and phonons. These diagonal states are eigenstates of the squeezing operator, then the photon-phonon mixed states are vacuum squeezed states. Even though, the independent photon and phonon states show no squeezing phenomena. The results are presented by calculating the squeezing parameters that are related to the uncertainties of the quadrature operators for the different quantum states \cite{Loudon2000}.

The paper starts in section 2 by introducing the quadratic Hamiltonian that represents the creation and annihilation of photon-phonon pairs. The daigonalization of the Hamiltonian is given in section 3 which yields photon-phonon entangled states. The squeezing properties of the mixed states appear in section 4 by calculating the Heisenberg uncertainties of the quadrature operators, where detail calculations are presented in the Appendix. Section 5 includes the conclusion.

\section{Interacting Photons and Phonons in Nanoscale Waveguides}

\begin{figure}
\includegraphics[width=0.8\linewidth]{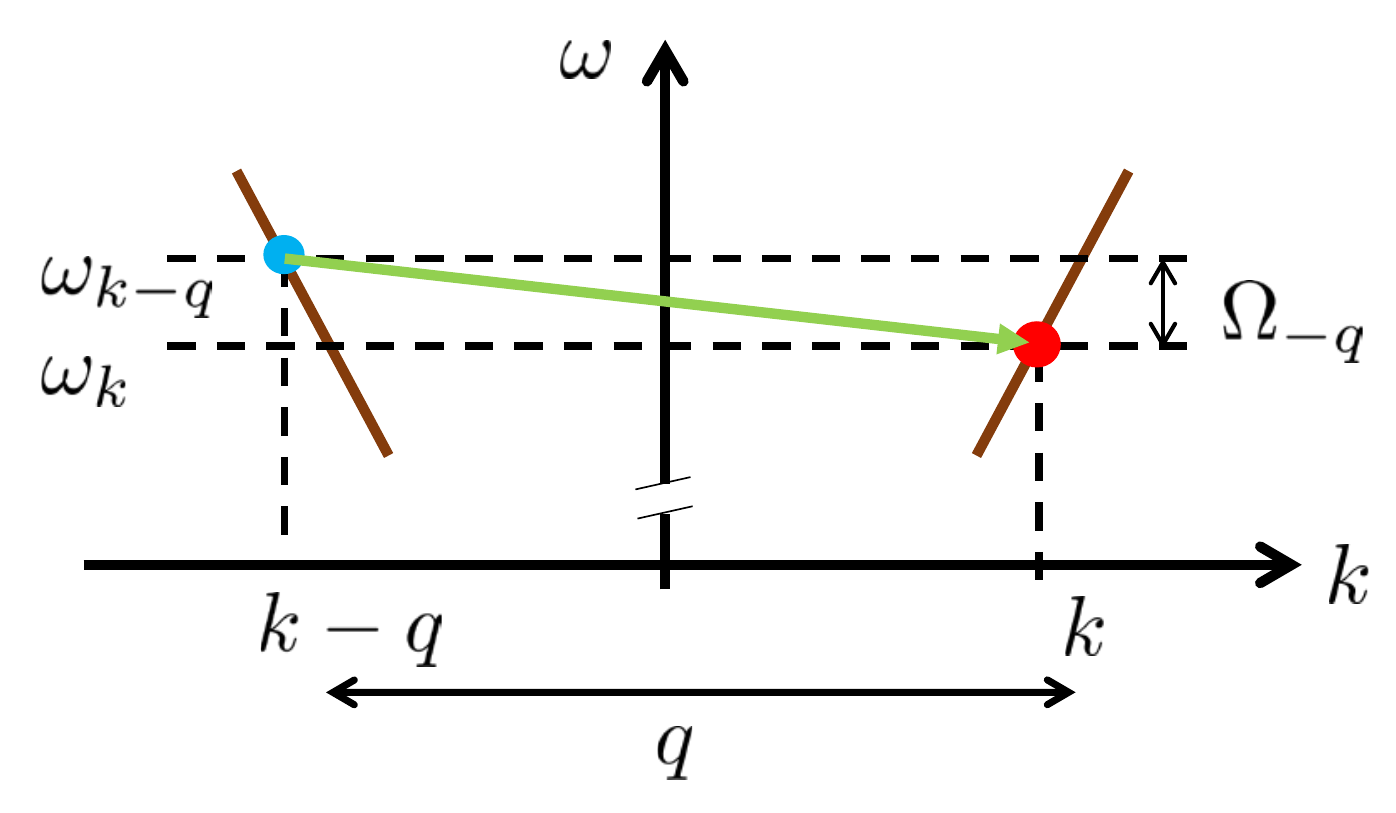}
\caption{Schematic plot of the photon dispersion, $\omega$ vs. $k$, for positive and negative wavenumbers. The phonon dispersion, $\Omega$ vs. $q$, is presented. The process of the scattering of a pump photon, $\omega_{k-q}$, into a signal photon, $\omega_{k}$, with the emission of a phonon, $\Omega_{-q}$, is plotted.}
\label{PhotPhonDis}
\end{figure}

We present first photons and phonons in nanoscale waveguides made of high contrast dielectric material. In our previous work \cite{Zoubi2016} we performed analytical calculations for obtaining the exact dispersions of both the electromagnetic field and the mechanical vibration modes inside a nanoscale waveguide of circular cross section made of silicon material (with refractive index of 3.45) that is localized in free space. The case of a nanoscale waveguide made of silicon with rectangular cross section is achieved numerically in \cite{Rakish2012}. The electromagnetic field can freely propagate along the waveguide
axis with almost continuous wavenumbers, for enough long waveguide, and is strongly confined in the transverse direction with discrete modes. Here we
assume a single active mode in the transverse direction. Exploiting translational symmetry, the wavenumber takes the values
$k=\frac{2\pi}{L}n$, where $L$ is the waveguide length with
$(n=0,\pm1,\pm2,\cdots,\pm\infty)$. Moreover, we concentrate in a region where light can propagate with almost
linear dispersion of an effective group velocity of $v_g$. Hence, the photon
dispersion is given by $\omega_k=\omega_0+v_gk$, around a given frequency  $\omega_0$ in the appropriate region \cite{Zoubi2016}. The photon dispersion
is schematically plotted in figure (\ref{PhotPhonDis}). The photon
Hamiltonian in momentum-space representation reads
\begin{equation}
H_{phot}=\sum_k\hbar\omega_k\ a_k^{\dagger}a_k,
\end{equation}
where $a_k$ and $a_k^{\dagger}$ are the creation and annihilation operators of
a photon with wavenumber $k$ for a given transverse mode and a fixed
polarization, with frequency $\omega_k$.

Next we consider the mechanical vibrations in the waveguide. In nanoscale
waveguides different types of vibrational modes can be excited, but here we consider only the propagating
acoustic phonons \cite{Zoubi2016}. The acoustic phonons have a linear
dispersion that is given by $\Omega_q=v_aq$, where $q$ is the acoustic phonon
wavenumber and $v_a$ is the sound velocity. The phonon dispersion is illustrated in
figure (\ref{PhotPhonDis}) for the acoustic mode. The phonon Hamiltonian reads
\begin{equation}
H_{phon}=\sum_{q}\hbar\Omega_{q}\ b_q^{\dagger}b_q,
\end{equation}
where $b_q$ and $b_q^{\dagger}$ are the creation and annihilation operators of
a phonon with wavenumber $q$ for an acoustic mode, of frequencies $\Omega_q$.

The SBS between photons and phonons is subjected to conservation of energy and
momentum. The SBS Hamiltonian is given by \cite{Zoubi2016}
\begin{equation}
H_{SBS}=\hbar\sum_{kq}\left(g_{kq}^{\ast}\ a_{k-q}^{\dagger}a_{k}b_{-q}+g_{kq}\ a_k^{\dagger}b_{-q}^{\dagger}a_{k-q}\right),
\end{equation}
where $g_{kq}$ is the SBS coupling parameter among the two photons of
wavenumbers $k$ and $k-q$ and a phonon of wavenumber $-q$. The coupling parameter can be in general $k$ and $q$ dependent, but we assume here the local
field approximation, that is $g_{kq}=g$. The exact dependent of the coupling parameter on the photon and phonon wavenumbers is given in \cite{Zoubi2016} for a nanoscale waveguide of $1$~cm length made of silicon with circular cross section of $500$~nm diameter. The first interaction term represents a process in which a photon, $(k,\omega_k)$, scatters into another photon, $(k-q,\omega_{k-q})$, with the absorption of a
phonon, $(-q,\Omega_{-q})$, under the conservation of energy,
$\omega_{k-q}-\omega_{k}=\Omega_{-q}$. The second term represent a photon, $(k-q,\omega_{k-q})$, that emits a phonon, $(-q,\Omega_{-q})$, and scatters into another photon, $(k,\omega_k)$, under the conservation of energy. The process is represented schematically in figure (\ref{PhotPhonDis}).

\begin{figure}
\includegraphics[width=0.8\linewidth]{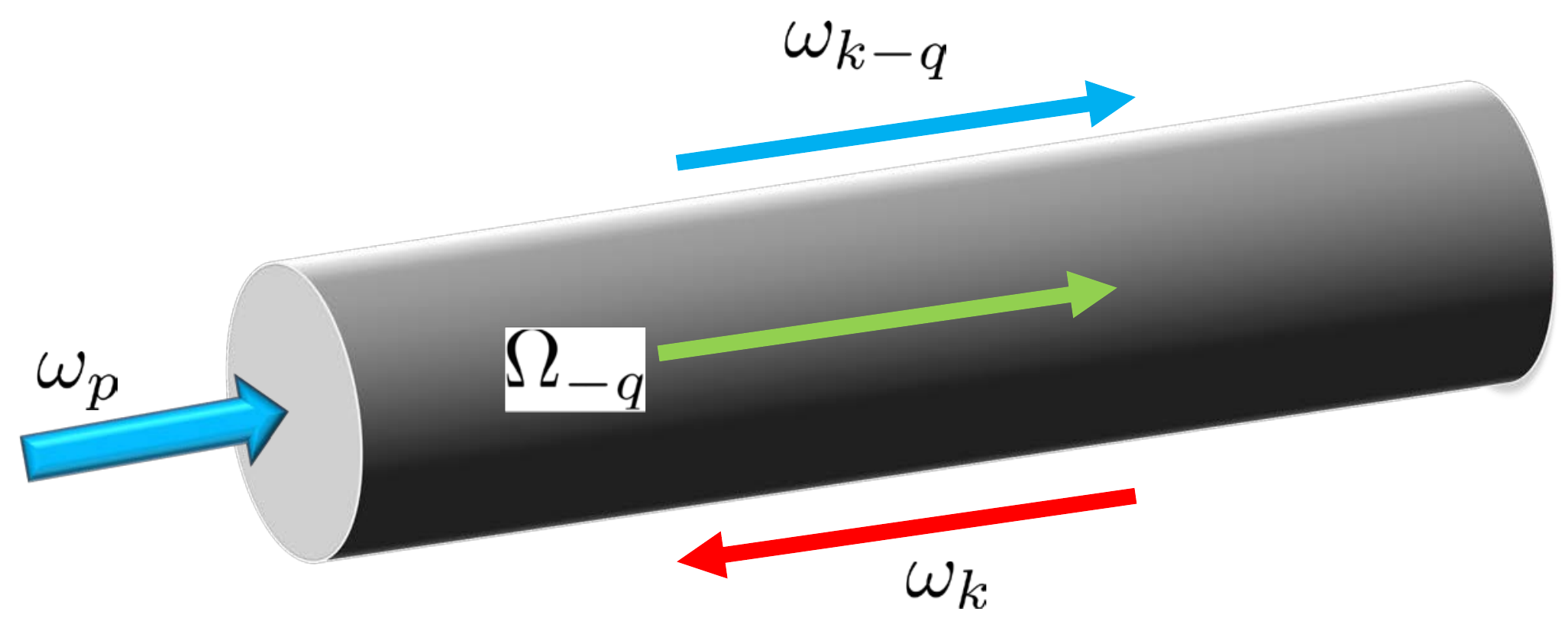}
\caption{The linear nanoscale waveguide in which the two sides are considered to be two effective mirrors, and which allow coupling among external radiation field and internal waveguide field. Inside the waveguide three fields are represented, the photons $\omega_k$, $\omega_{k-q}$ and the phonon $\Omega_{-q}$. An external pump field also appears, $\omega_p$, on the left side.}
\label{Linear}
\end{figure}

In order to excite the waveguide photons one need to couple the internal field to the external radiation field in using the input-output formalism \cite{Walls2008,Gardiner2010}. We consider the linear waveguide of a finite length and we assume two effective mirrors at the two edges, as seen in figure (\ref{Linear}). The mirrors serve us to couple
the external field with the internal waveguide field, and also photons can leak
outside the waveguide through the mirrors. The input and output external fields
of wavenumber $k$ are given by $c_{k}^{in}$ and
$c_{k}^{out}$, and which are related to the waveguide mode $k$ by the boundary condition $c_{k}^{in}+c_{k}^{out}=\sqrt{u_{k}}a_k$, where $u_{k}$ is the effective coupling parameter between the internal and
external fields of wavenumber $k$. In the following we assume identical mirrors and the
coupling is independent of the wavenumber where
$u_{k}=u$. For example, in tapered nanofibers made of silica the tapered zones can be considered as effective mirrors \cite{Vetsch2010,Goban2012}. For nanophotonic waveguides, e.g. made of silicon, grating couplers are used to couple the internal-external fields that considered as effective mirrors \cite{Taillaert2006}.

We concentrate in the case of three fields inside the waveguide, a signal photon, $(k,\omega_k)$, a pump photon, $(k-q,\omega_{k-q})$, and an acoustic phonon, $(-q,\Omega_{-q})$. Therefore, we limit our discussion to the Hamiltonian
\begin{eqnarray} \label{QuadHam}
H&=&\hbar\omega_{k}\ a_k^{\dagger}a_k+\hbar\omega_{k-q}\ a_{k-q}^{\dagger}a_{k-q}+\hbar\Omega_{-q}\ b^{\dagger}_{-q}b_{-q} \nonumber \\
&+&\hbar g^{\ast}\ a^{\dagger}_{k-q}a_{k}b_{-q}+\hbar g\ b_{-q}^{\dagger}a_k^{\dagger}a_{k-q}.
\end{eqnarray}

The first step is to bring the Hamiltonian into a quadratic form that includes
effective coupling between signal
photons and phonons that induced by SBS in the presence of the pump field. This procedure can be achieved in assuming the strong pump
field to be a classical one, where the obtained effective photon-phonon coupling is found to be related on the
pump field intensity. 

We apply an external control field, $(\omega_p,k-q)$, that is described by $c_{k-q}^{in}$, in order to excite the waveguide pump field at frequency $\omega_{k-q}$, where the fields are close to resonance $\omega_{k-q}\approx\omega_p$. From conservation of energy and momentum, and due to SBS within the waveguide, a pump photon is scattered into a signal photon of $\omega_k$ assisted by the emission of an acoustic phonon of $\Omega_{-q}$, where $\omega_{k-q}\approx\omega_{k}+\Omega_{-q}$ with the required phase-matching.

Let us concentrate now in the pump field at wavenumber $k-q$. The input pump field is represented by $c_{k-q}^{in}$. The pump field is enough strong in order to neglect any change in it due to SBS. Hence, using the input-output formalism \cite{Gardiner2010}, we get the equation of motion for the pump operator
\begin{equation}
\frac{d}{dt}\tilde{a}_{k-q}\approx-i\Delta_{k-q}\ \tilde{a}_{k-q}+\sqrt{u}\ \tilde{c}_{k-q}^{in},
\end{equation}
where $\Delta_{k-q}=\omega_{k-q}-\omega_p-i\left(u+\frac{\gamma}{2}\right)$, and we defined $a_{k-q}=\tilde{a}_{k-q}e^{-i\omega_pt}$, with $c_{k-q}^{in}=\tilde{c}_{k-q}^{in}e^{-i\omega_pt}$. The photon direct damping into free space is included phenomenologically through the damping rate $\gamma$. At steady state we have $\frac{d}{dt}\tilde{a}_{k-q}=0$. Then, for the pump field we get
\begin{equation} \label{pump}
a_{k-q}=\frac{\sqrt{u}}{i\Delta_{k-q}}c_{k-q}^{in}.
\end{equation}
We define $\hat{n}_{k-q}=a_{k-q}^{\dagger}a_{k-q}$, where $\hat{n}_{k-q}=\frac{u}{|\Delta_{k-q}|^2}\hat{n}_{k-q}^{in}$, with $\hat{n}_{k-q}^{in}=c_{k-q}^{in\dagger}c_{k-q}^{in}$. The average
value is $n_{k-q}=\langle a_{k-q}^{\dagger}a_{k-q}\rangle$, then $n_{k-q}=\frac{u}{|\Delta_{k-q}|^2}n_{k-q}^{in}$, with $n_{k-q}^{in}=\langle c_{k-q}^{in\dagger}c_{k-q}^{in}\rangle$. Note that $n_{k-q}^{in}$ has a unit of the number of photons per second.

\section{Photon-Phonon Bogoliubov Modes}

We start from the above Hamiltonian (\ref{QuadHam}) that includes only the three appropriate waveguide fields. The system can be linearized by assuming a classical pump
field, where the Hamiltonian casts into a quadratic one. We use the
previous results, where the pump field operator, at steady state in using
the input-output formalism, is given by equation (\ref{pump}). In the
classical limit we have $\lambda_{k-q}=\frac{\sqrt{u}}{i\Delta_{k-q}}\mu_{k-q}$, where $\lambda_{k-q}=\langle a_{k-q}\rangle$, and
$\mu_{k-q}=\langle c_{k-q}^{in}\rangle$. We achieve the quadratic Hamiltonian
\begin{eqnarray}
H&=&\hbar\omega_{k}\ a_k^{\dagger}a_k+\hbar\Omega_{-q}\ b^{\dagger}_{-q}b_{-q} \nonumber \\
&+&\hbar g^{\ast}\lambda_{k-q}^{\ast}\ b_{-q}a_k+\hbar g\lambda_{k-q}\ b_{-q}^{\dagger}a_k^{\dagger}.
\end{eqnarray}
We move to a rotating frame that oscillates with the pump waveguide frequency $\omega_{k-q}$, to
get the Hamiltonian
\begin{eqnarray}
H&=&\hbar\bar{\omega}_{kq}\ a^{\dagger}_ka_k+\hbar\Omega_{-q}\ b^{\dagger}_{-q}b_{-q} \nonumber \\
&+&\hbar f^{\ast}_{k-q}\ b_{-q}a_k+\hbar f_{k-q}\ b_{-q}^{\dagger}a^{\dagger}_k,
\end{eqnarray}
where $\bar{\omega}_{kq}=\omega_{k-q}-\omega_{k}$, with the effective coupling
\begin{equation}
f_{k-q}=\frac{g\sqrt{u}}{i\Delta_{k-q}}\mu_{k-q}.
\end{equation}
We treat the case of resonance external-internal fields, where $\omega_k=\omega_s$ and $\omega_{k-q}=\omega_p$, then $\bar{\omega}_{kq}=\omega_p-\omega_s$, with $i\Delta_{k-q}=u+\frac{\gamma}{2}$.

Here we concentrate in the case of fixed wavenumbers, $k$ and $q$. For further simplification we use the replacement $a_k\rightarrow \hat{a}$ and $b_{-q}\rightarrow \hat{b}$, with $f_{k-q}\rightarrow f$, $\Omega_{-q}\rightarrow \Omega$, and $\bar{\omega}_{kq}\rightarrow \omega$, where $\omega=\omega_p-\omega_s$. Hence we can write, in assuming real $f$,
\begin{equation}
H=\hbar\omega\ \hat{a}^{\dagger}\hat{a}+\hbar\Omega\ \hat{b}^{\dagger}\hat{b}+\hbar f\left( \hat{b}\hat{a}+\hat{b}^{\dagger}\hat{a}^{\dagger}\right).
\end{equation}
The Hamiltonian can be easily diagonalized using the Bogoliubov transformation \cite{Fetter1971}
\begin{eqnarray}\label{TRANS}
\hat{\alpha}&=&\cosh r\ \hat{a}+\sinh r\ \hat{b}^{\dagger}, \nonumber \\
\hat{\beta}&=&\cosh r\ \hat{b}+\sinh r\ \hat{a}^{\dagger},
\end{eqnarray}
with the inverse transformation
\begin{eqnarray}
\hat{a}&=&\cosh r\ \hat{\alpha}-\sinh r\ \hat{\beta}^{\dagger}, \nonumber \\
\hat{b}&=&\cosh r\ \hat{\beta}-\sinh r\ \hat{\alpha}^{\dagger}.
\end{eqnarray}
The fact that $\hat{a}$ and $\hat{b}$ obey boson commutation relations, $[\hat{a},\hat{a}^{\dagger}]=[\hat{b},\hat{b}^{\dagger}]=1$ and $[\hat{a},\hat{b}^{\dagger}]=0$, lead to $[\hat{\alpha},\hat{\alpha}^{\dagger}]=[\hat{\beta},\hat{\beta}^{\dagger}]=1$ and $[\hat{\alpha},\hat{\beta}^{\dagger}]=0$, hence $\cosh^2 r-\sinh^2 r=1$. Substitution in the Hamiltonian gives
\begin{eqnarray}
H&=&\hbar\left[(\omega+\Omega)\sinh^2 r-2f\cosh r\sinh r\right]\hat{\mathbb{I}} \nonumber \\
&+&\hbar\left[\omega\cosh^2 r+\Omega\sinh^2 r-2f\cosh r\sinh r\right]\hat{\alpha}^{\dagger}\hat{\alpha} \nonumber \\
&+&\hbar\left[\Omega\cosh^2 r+\omega\sinh^2 r-2f\cosh r\sinh r\right]\hat{\beta}^{\dagger}\hat{\beta} \nonumber \\
&-&\hbar\left[(\omega+\Omega)\cosh r\sinh r-f(\cosh^2 r+\sinh^2 r)\right] \nonumber \\
&\times&\left(\hat{\alpha}\hat{\beta}+\hat{\alpha}^{\dagger}\hat{\beta}^{\dagger}\right),
\end{eqnarray}
where $\hat{\mathbb{I}}$ is a unit operator. We choose $(\omega+\Omega)\cosh r\sinh r=f(\cosh^2 r+\sinh^2 r)$, to get the diagonal Hamiltonian
\begin{equation}
H=\hbar\omega_0\ \hat{\mathbb{I}}+\hbar\omega_{\alpha}\ \hat{\alpha}^{\dagger}\hat{\alpha}+\hbar\omega_{\beta}\ \hat{\beta}^{\dagger}\hat{\beta},
\end{equation}
where
\begin{eqnarray}
\omega_0&=&(\omega+\Omega)\sinh^2 r-2f\cosh r\sinh r, \nonumber \\
\omega_{\alpha}&=&\omega\cosh^2 r+\Omega\sinh^2 r-2f\cosh r\sinh r, \nonumber \\
\omega_{\beta}&=&\Omega\cosh^2 r+\omega\sinh^2 r-2f\cosh r\sinh r.
\end{eqnarray}
The calculation yields
\begin{equation}
\cosh^2 r=\frac{\bar{\omega}+\Delta}{2\Delta},\ \sinh^2 r=\frac{\bar{\omega}-\Delta}{2\Delta},
\end{equation}
and $\cosh r\sinh r=\frac{f}{2\Delta}$, where we defined $\Delta^2=\bar{\omega}^2-f^2$, with $\bar{\omega}=\frac{\omega+\Omega}{2}$. We obtain
\begin{equation}
\omega_{\alpha}=\Delta+\delta,\ \omega_{\beta}=\Delta-\delta,
\end{equation}
and $\omega_0=\Delta-\bar{\omega}$,
where $\delta=\frac{\omega-\Omega}{2}$.

The ground state $|vac\rangle$ has energy $\hbar\omega_0$. We get two collective Bogoliubov modes with energies $\hbar\omega_{\alpha}$ and $\hbar\omega_{\beta}$, which are represented by the boson operators $\hat{\alpha}$ and $\hat{\beta}$, respectively. The creation and annihilation of a collective excitation involves creation and annihilation of a photon and a phonon in the presence of an external pump field. The collective states are entangled states of photons and phonons, as discussed in the next section. The stability requirement implies the condition $\bar{\omega}>f$, which can be easily achieved in nanoscale waveguides.

\section{Photon-Phonon Squeezed States} 

Here we present a deep study of the collective states obtained from the creation and annihilation of a pair of photons and phonons in the presence of a pump field in nanophotonic waveguides. We concentrate in the entanglement between photons and phonons and emphasize the squeezing in the uncertainties of the quadrature operators of the photon-phonon mixed states. In the light of the previous results we define the squeezed operator \cite{Loudon2000}
\begin{equation}
\hat{S}(r)=e^{r\left(\hat{a}^{\dagger}\hat{b}^{\dagger}-\hat{a}\hat{b}\right)},
\end{equation}
which can be written as
\begin{equation}
\hat{S}(r)=e^{\tanh r\ \hat{a}^{\dagger}\hat{b}^{\dagger}}e^{-\ln\cosh r\left(\hat{a}^{\dagger}\hat{a}+\hat{b}^{\dagger}\hat{b}+\hat{\mathbb{I}}\right)}e^{-\tanh r\ \hat{a}\hat{b}}.
\end{equation}
We define the photon and phonon Fock's state $|n_{phot},n_{phon}\rangle$. The vacuum state is $|vac\rangle=|0_{phot},0_{phon}\rangle$, where $\hat{a}|0_{phot},0_{phon}\rangle=\hat{b}|0_{phot},0_{phon}\rangle=0$. We apply the $\hat{S}(r)$ operator to the photon and phonon operators, by using
\begin{equation} 
\hat{\alpha}=\hat{S}^{\dagger}(r)\hat{a}\hat{S}(r),\ \hat{\beta}=\hat{S}^{\dagger}(r)\hat{b}\hat{S}(r),
\end{equation}
which yield exactly equations (\ref{TRANS}).

Now, we apply the $\hat{S}(r)$ operator on the vacuum state. We define
\begin{equation}
|r\rangle=\hat{S}(r)|vac\rangle,
\end{equation}
that yields
\begin{equation}
|r\rangle=\frac{1}{\cosh r}e^{\tanh r\ \hat{a}^{\dagger}\hat{b}^{\dagger}}|0_{phot},0_{phon}\rangle.
\end{equation}
In terms of photon and phonon Fock states we have
\begin{equation}
|r\rangle=\frac{1}{\cosh r}\sum_{n=0}^{\infty}\tanh^nr|n,n\rangle,
\end{equation}
where $|n,n\rangle=|n_{phot},n_{phon}\rangle$. The state represents entanglement between photons and phonons. Hence, if $n$ photons exist in the system then for sure exists $n$ phonons. The probability of finding $n$ photons and $n$ phonons is given by
\begin{equation}
P_{n}=\frac{\tanh^{2n} r}{\cosh^2 r}.
\end{equation}
In the limit of $\tanh r\ll1$, we can expand the $|r\rangle$ state in terms of Fock states as
\begin{equation} \label{Ent}
|r\rangle\approx\left(|0,0\rangle+r\ |1,1\rangle+\cdots\right).
\end{equation}
Keeping the first two terms gives photon-phonon entangled state of the Bell's state type. The limit of $\tanh r\ll1$ (or $r\ll1$) can be achieved at $\bar{\omega}\gg f$, which is consistent with the stability condition $\bar{\omega}>f$. Our main concern now is to show that the $|r\rangle$ state represents a vacuum squeezed state for a mix of photons and phonons. Moreover, the above state can be used as a source of single phonons, where once a single photon is observed in the system then for sure a single phonon exists \cite{Riedinger2016}.

For a nanoscale cylindrical waveguide of $500$~nm diameter one can consider a phonon with $\Omega=10$~GHz frequency, and the photon-phonon coupling parameter is $g=1$~MHz \cite{Zoubi2016}. The input-output coupling parameter can be of $u=1$~MHz and the photon damping rate is $\gamma=10$~mHz. At resonance, that is $\delta=0$, one can choose the pump intensity such that the effective coupling parameter is $f=1$~GHz. Here the pump intensity is of $n^{in}=10^{12}$~sec$^{-1}$. In this case we have $\cosh^2r\approx1.0025$ and $\sinh^2r\approx0.0025$ where $\tanh r\approx0.05$ and then $r\approx 0.05$. We obtain $P_0\approx 0.9975$, $P_1\approx 0.0025$ and $P_2\approx 6.25\times 10^{-6}$. As $r\ll 1$ hence the above expansion of equation (\ref{Ent}) holds. Note that for resonance the limit of $r\ll1$ holds at $\Omega\gg f$ as here $\bar{\omega}=\omega=\Omega$.

\subsection{Independent Photons and Phonons}

We start by defining the quadrature operators for both the photons and phonons \cite{Loudon2000}. We have for the photons
\begin{equation}
\hat{X}_a=\frac{\hat{a}+\hat{a}^{\dagger}}{\sqrt{2}},\ \hat{Y}_a=\frac{\hat{a}-\hat{a}^{\dagger}}{i\sqrt{2}},
\end{equation}
with the uncertainties
\begin{equation}
(\Delta\hat{X}_a)^2=\langle\hat{X}_a^2\rangle-\langle\hat{X}_a\rangle^2,\ (\Delta\hat{Y}_a)^2=\langle\hat{Y}_a^2\rangle-\langle\hat{Y}_a\rangle^2,
\end{equation}
and for the phonons
\begin{equation}
\hat{X}_b=\frac{\hat{b}+\hat{b}^{\dagger}}{\sqrt{2}},\ \hat{Y}_b=\frac{\hat{b}-\hat{b}^{\dagger}}{i\sqrt{2}},
\end{equation}
with the uncertainties
\begin{equation}
(\Delta\hat{X}_b)^2=\langle\hat{X}_b^2\rangle-\langle\hat{X}_b\rangle^2,\ (\Delta\hat{Y}_b)^2=\langle\hat{Y}_b^2\rangle-\langle\hat{Y}_b\rangle^2.
\end{equation}
We calculate the expectation values of the photon and phonon quadrature operators in the state $|r\rangle$. In the Appendix we give detail calculations. We get $\langle\hat{X}_a\rangle=\langle\hat{Y}_a\rangle=\langle\hat{X}_b\rangle=\langle\hat{Y}_b\rangle=0$. Furthermore, we obtain 
\begin{equation}
\langle\hat{X}_a^2\rangle=\langle\hat{Y}_a^2\rangle=\langle\hat{X}_b^2\rangle=\langle\hat{Y}_b^2\rangle=\frac{1}{2}\cosh 2r.
\end{equation}
The uncertainties are
\begin{eqnarray}
\Delta\hat{X}_a=\sqrt{\langle\hat{X}_a^2\rangle}&,&\Delta\hat{Y}_a=\sqrt{\langle\hat{Y}_a^2\rangle}, \nonumber \\
\Delta\hat{X}_b=\sqrt{\langle\hat{X}_b^2\rangle}&,&\Delta\hat{Y}_b=\sqrt{\langle\hat{Y}_b^2\rangle}.
\end{eqnarray}
All uncertainties are equal, where
\begin{equation}
\Delta\hat{X}_a=\Delta\hat{Y}_a=\Delta\hat{X}_b=\Delta\hat{Y}_b=\frac{1}{\sqrt{2}}\sqrt{\cosh^2r+\sinh^2r}.
\end{equation}
The Heisenberg uncertainty relations are
\begin{equation}
\Delta\hat{X}_a\Delta\hat{Y}_a=\Delta\hat{X}_b\Delta\hat{Y}_b=\frac{1}{2}+\sinh^2r,
\end{equation}
which is larger than $1/2$ (equals $1/2$ at $r=0$).

The squeezing parameter is defined by
\begin{eqnarray}
S_a^X=\left(\Delta\hat{X}_a\right)^2-\frac{1}{2}&,&S_a^Y=\left(\Delta\hat{Y}_a\right)^2-\frac{1}{2}, \nonumber \\
S_b^X=\left(\Delta\hat{X}_b\right)^2-\frac{1}{2}&,&S_b^Y=\left(\Delta\hat{Y}_b\right)^2-\frac{1}{2}.
\end{eqnarray}
Here we get the same squeezing parameter for all quadrature of both photons and phonons
\begin{equation}
S_a^X=S_a^Y=S_b^X=S_b^Y=\sinh^2r.
\end{equation}
The squeezing parameters are positive, and then no squeezing appears neither for the photon or the phonon independent modes. For example, using the previous numbers we get $S_a^X=S_a^Y=S_b^X=S_b^Y\approx +0.0025$.

\subsection{Mixed States of Photons and Phonons}

Let us define the photon-phonon mixed operators
\begin{equation}
\hat{c}=\frac{\hat{a}-\hat{b}}{\sqrt{2}},\ \hat{d}=\frac{\hat{a}+\hat{b}}{\sqrt{2}},
\end{equation}
where
\begin{equation}
[\hat{c},\hat{c}^{\dagger}]=[\hat{d},\hat{d}^{\dagger}]=1,\ [\hat{c},\hat{d}^{\dagger}]=0.
\end{equation}
We define the related quadrature operators
\begin{equation}
\hat{X}_c=\frac{\hat{c}+\hat{c}^{\dagger}}{\sqrt{2}},\ \hat{Y}_c=\frac{\hat{c}-\hat{c}^{\dagger}}{i\sqrt{2}},
\end{equation}
with the uncertainties
\begin{equation}
(\Delta\hat{X}_c)^2=\langle\hat{X}_c^2\rangle-\langle\hat{X}_c\rangle^2,\ (\Delta\hat{Y}_c)^2=\langle\hat{Y}_c^2\rangle-\langle\hat{Y}_c\rangle^2,
\end{equation}
and
\begin{equation}
\hat{X}_d=\frac{\hat{d}+\hat{d}^{\dagger}}{\sqrt{2}},\ \hat{Y}_d=\frac{\hat{d}-\hat{d}^{\dagger}}{i\sqrt{2}},
\end{equation}
with the uncertainties
\begin{equation}
(\Delta\hat{X}_d)^2=\langle\hat{X}_d^2\rangle-\langle\hat{X}_d\rangle^2,\ (\Delta\hat{Y}_d)^2=\langle\hat{Y}_d^2\rangle-\langle\hat{Y}_d\rangle^2.
\end{equation}
In the Appendix we calculate the expectation values, to get $\langle\hat{X}_c\rangle=\langle\hat{Y}_c\rangle=\langle\hat{X}_d\rangle=\langle\hat{Y}_d\rangle=0$, and
\begin{eqnarray}
\langle\hat{X}_c^2\rangle=\frac{1}{2}e^{-2r}&,&\langle\hat{Y}_c^2\rangle=\frac{1}{2}e^{2r}, \nonumber \\
\langle\hat{X}_d^2\rangle=\frac{1}{2}e^{2r}&,&\langle\hat{Y}_d^2\rangle=\frac{1}{2}e^{-2r}.
\end{eqnarray}
Then the uncertainties are
\begin{eqnarray}
\Delta\hat{X}_c&=&\sqrt{\langle\hat{X}_c^2\rangle}=\frac{1}{\sqrt{2}}e^{-r}, \nonumber  \\
\Delta\hat{Y}_c&=&\sqrt{\langle\hat{Y}_c^2\rangle}=\frac{1}{\sqrt{2}}e^{r}, \nonumber \\
\Delta\hat{X}_d&=&\sqrt{\langle\hat{X}_d^2\rangle}=\frac{1}{\sqrt{2}}e^{r}, \nonumber \\
\Delta\hat{Y}_d&=&\sqrt{\langle\hat{Y}_d^2\rangle}=\frac{1}{\sqrt{2}}e^{-r}.
\end{eqnarray}
The Heisenberg uncertainties are, as expected,
\begin{equation}
\Delta\hat{X}_c\Delta\hat{Y}_c=\Delta\hat{X}_d\Delta\hat{Y}_d=\frac{1}{2}.
\end{equation}

Now the squeezed parameters are defined by
\begin{eqnarray}
S_c^X=\left(\Delta\hat{X}_c\right)^2-\frac{1}{2}&,&S_c^Y=\left(\Delta\hat{Y}_c\right)^2-\frac{1}{2}, \nonumber \\
S_d^X=\left(\Delta\hat{X}_d\right)^2-\frac{1}{2}&,&S_d^Y=\left(\Delta\hat{Y}_d\right)^2-\frac{1}{2}.
\end{eqnarray}
We obtain
\begin{equation}
S_c^X=S_d^Y=\frac{1}{2}\left(e^{-2r}-1\right),\ S_c^Y=S_d^X=\frac{1}{2}\left(e^{2r}-1\right).
\end{equation}
For example, using the previous numbers, where $r\approx 0.05$, we get $S_c^X=S_d^Y\approx -0.0475$ and $S_c^Y=S_d^X\approx +0.0525$. As the squeezed parameters can be negative then the mixed states are non-classical and show squeezing phenomena.

The effect of thermal phonons is critical for achieving squeezed states and for a source of single phonons. In typical nanoscale waveguides we have for the phonon damping rate $\Gamma = 1\ MHz$ \cite{VanLaer2017,Safavi2019}. For phonon frequency of $\Omega=10$~GHz, we get $Q=10^4$. At temperature of
$T=200$~mK, the average number of thermal phonons is $\bar{n}\approx0.1$.

\section{Conclusions}

Nanoscale waveguides are an ideal platform for manipulating propagating photons and phonons with strong mutual coupling. By limiting the discussion to two light fields and a single sound field, we obtained a squeezing type Hamiltonian by choosing one of the light fields to be strong and treated classically. The Hamiltonian represents creation and annihilation of photon-phonon pairs that subjected to conservation of energy and momentum with the aid of the pump field. Here the frequency and intensity of the external pump field serve as control parameters for the photon-phonon coupling and their resonance. The Hamiltonian can be easily diagonalized using the known Bogoliubov transformation to get diagonal eigenstates that are a coherent mix of photons and phonons.

The diagonal eigenstates are shown to be also eigenstates of the squeezing operator and they form entanglement between the photon and phonon Fock states. At small squeezing number the observation of a single photon yields for sure a single phonon in the system, which used as an efficient source of single phonons at relatively low temperature. The calculation of the uncertainties of the quadrature operators for the independent photons and phonons show no squeezing phenomena. But, the uncertainties of the quadrature operators for a coherent mix of the phonons and photons show squeezing properties. The negativity of the squeezing parameter is a measure for non-classical behavior.

The generation of entangled states of photons and phonons with squeezing properties is of importance for fundamental physics and applications. The hybridization of photons and phonons inside the same setup combines the properties of both light and sound modes. This fact is useful for the physical implementation of photons and phonons in quantum information processing. The result can provide a controllable source of single phonons. The setup is a solid state component and can be easily integrated into all-optical on-ship platform, and can be realized in using, e.g., a nanowire made of silicon or a silica tapered nanfiber.

\section*{Acknowledgment}

The work was supported by the Council for Higher Education in Israel via the Maa'of Grant.

\section*{Appendix}

In this Appendix we calculate the expectation values of the quadrature operators that are needed in order to get the uncertainties in the main text.

We start by calculating the expectation values for the independent photon and phonon states. We have
\begin{eqnarray}
\langle\hat{X}_a\rangle&=&\frac{1}{\sqrt{2}}\langle r|\left(\hat{a}+\hat{a}^{\dagger}\right)|r\rangle \nonumber \\
&=&\frac{1}{\sqrt{2}}\langle vac|\hat{S}^{\dagger}\left(\hat{a}+\hat{a}^{\dagger}\right)\hat{S}|vac\rangle \nonumber \\
&=&\frac{1}{\sqrt{2}}\langle vac|\left(\hat{\alpha}+\hat{\alpha}^{\dagger}\right)|vac\rangle.
\end{eqnarray}
From (\ref{TRANS}) and using the fact that $\langle vac|\hat{a}|vac\rangle=\langle vac|\hat{a}^{\dagger}|vac\rangle=\langle vac|\hat{b}|vac\rangle=\langle vac|\hat{b}^{\dagger}|vac\rangle=0$, we have $\langle vac|\hat{\alpha}|vac\rangle=\langle vac|\hat{\alpha}^{\dagger}|vac\rangle=\langle vac|\hat{\beta}|vac\rangle=\langle vac|\hat{\beta}^{\dagger}|vac\rangle=0$. We get $\langle\hat{X}_a\rangle=0$, and similar calculations lead to $\langle\hat{Y}_a\rangle=\langle\hat{X}_b\rangle=\langle\hat{Y}_b\rangle=0$.

Now we calculate the photon quadratic terms
\begin{eqnarray}
\langle\hat{X}_a^2\rangle&=&\frac{1}{2}\langle r|\left(\hat{a}+\hat{a}^{\dagger}\right)\left(\hat{a}+\hat{a}^{\dagger}\right)|r\rangle \nonumber \\
&=&\frac{1}{2}\langle vac|\hat{S}^{\dagger}\left(\hat{a}^2+(\hat{a}^{\dagger})^2+\hat{a}^{\dagger}\hat{a}+\hat{a}\hat{a}^{\dagger}\right)\hat{S}|vac\rangle \nonumber \\
&=&\frac{1}{2}\langle vac|\left(\hat{\alpha}^2+(\hat{\alpha}^{\dagger})^2+2\hat{\alpha}^{\dagger}\hat{\alpha}+1\right)|vac\rangle,
\end{eqnarray}
and
\begin{equation}
\langle\hat{Y}_a^2\rangle=\frac{1}{2}\langle vac|\left(2\hat{\alpha}^{\dagger}\hat{\alpha}+1-\hat{\alpha}^2-(\hat{\alpha}^{\dagger})^2\right)|vac\rangle.
\end{equation}
For the phonon we obtain
\begin{eqnarray}
\langle\hat{X}_b^2\rangle&=&\frac{1}{2}\langle vac|\left(\hat{\beta}^2+(\hat{\beta}^{\dagger})^2+2\hat{\beta}^{\dagger}\hat{\beta}+1\right)|vac\rangle, \nonumber \\
\langle\hat{Y}_b^2\rangle&=&\frac{1}{2}\langle vac|\left(2\hat{\beta}^{\dagger}\hat{\beta}+1-\hat{\beta}^2-(\hat{\beta}^{\dagger})^2\right)|vac\rangle.
\end{eqnarray}
On the other hand, using (\ref{TRANS}), we have
\begin{eqnarray}
\hat{\alpha}^{\dagger}\hat{\alpha}&=&\cosh^2r\ \hat{a}^{\dagger}\hat{a}+\sinh^2r\ \hat{b}\hat{b}^{\dagger} \nonumber \\
&+&\cosh r\sinh r\left(\hat{a}\hat{b}+\hat{a}^{\dagger}\hat{b}^{\dagger}\right), \nonumber \\
\hat{\alpha}^2&=&\cosh^2r\ \hat{a}^2+\sinh^2r\ \left(\hat{b}^{\dagger}\right)^2 \nonumber \\
&+&2\cosh r\sinh r\ \hat{b}^{\dagger}\hat{a},
\end{eqnarray}
and
\begin{eqnarray}
\hat{\beta}^{\dagger}\hat{\beta}&=&\cosh^2r\ \hat{b}^{\dagger}\hat{b}+\sinh^2r\ \hat{a}\hat{a}^{\dagger} \nonumber \\
&+&\cosh r\sinh r\left(\hat{a}\hat{b}+\hat{a}^{\dagger}\hat{b}^{\dagger}\right), \nonumber \\
\hat{\beta}^2&=&\cosh^2r\ \hat{b}^2+\sinh^2r\ \left(\hat{a}^{\dagger}\right)^2 \nonumber \\
&+&2\cosh r\sinh r\ \hat{a}^{\dagger}\hat{b},
\end{eqnarray}
in using 
\begin{eqnarray}\label{Ex3}
\langle vac|\hat{a}^{\dagger}\hat{a}|vac\rangle=\langle vac|\hat{b}^{\dagger}\hat{b}|vac\rangle&=& \nonumber \\
\langle vac|\hat{a}\hat{b}|vac\rangle=\langle vac|\hat{a}^{\dagger}\hat{b}|vac\rangle&=&0,
\end{eqnarray}
and also
\begin{equation}\label{Ex4}
\langle vac|\hat{a}^2|vac\rangle=\langle vac|\hat{b}^2|vac\rangle=0,
\end{equation}
with
\begin{equation}\label{Ex5}
\langle vac|\hat{a}\hat{a}^{\dagger}|vac\rangle=\langle vac|\hat{b}\hat{b}^{\dagger}|vac\rangle=1,
\end{equation}
that yield
\begin{equation} \label{EX1}
\langle vac|\hat{\alpha}^{\dagger}\hat{\alpha}|vac\rangle=\langle vac|\hat{\beta}^{\dagger}\hat{\beta}|vac\rangle=\sinh^2r,
\end{equation}
and
\begin{equation} \label{EX2}
\langle vac|\hat{\alpha}^2|vac\rangle=\langle vac|\hat{\beta}^2|vac\rangle=0.
\end{equation}
We get
\begin{equation}
\langle\hat{X}_a^2\rangle=\langle\hat{Y}_a^2\rangle=\langle\hat{X}_b^2\rangle=\langle\hat{Y}_b^2\rangle=\frac{1}{2}\left(1+2\sinh^2r\right).
\end{equation}

Next we calculate the expectation values for the mixed photon and phonon states. We have
\begin{eqnarray}
\langle\hat{X}_c\rangle&=&\frac{1}{\sqrt{2}}\langle r|\left(\hat{c}+\hat{c}^{\dagger}\right)|r\rangle=\frac{1}{2}\langle r|\left(\hat{a}+\hat{a}^{\dagger}-\hat{b}-\hat{b}^{\dagger}\right)|r\rangle, \nonumber \\
&=&\frac{1}{2}\langle vac|\left(\hat{\alpha}+\hat{\alpha}^{\dagger}-\hat{\beta}-\hat{\beta}^{\dagger}\right)|vac\rangle,
\end{eqnarray}
that gives, using previous results, $\langle\hat{X}_c\rangle=0$, and similar calculations yields $\langle\hat{Y}_c\rangle=\langle\hat{X}_d\rangle=\langle\hat{Y}_d\rangle=0$. The quadratic operators are
\begin{eqnarray}
\langle\hat{X}_c^2\rangle&=&\frac{1}{2}\langle r|\left(\hat{c}^2+(\hat{c}^{\dagger})^2+2\hat{c}^{\dagger}\hat{c}+1\right)|r\rangle, \nonumber \\
\langle\hat{Y}_c^2\rangle&=&-\frac{1}{2}\langle r|\left(\hat{c}^2+(\hat{c}^{\dagger})^2-2\hat{c}^{\dagger}\hat{c}-1\right)|r\rangle, \nonumber \\
\langle\hat{X}_d^2\rangle&=&\frac{1}{2}\langle r|\left(\hat{d}^2+(\hat{d}^{\dagger})^2+2\hat{d}^{\dagger}\hat{d}+1\right)|r\rangle, \nonumber \\
\langle\hat{Y}_d^2\rangle&=&-\frac{1}{2}\langle r|\left(\hat{d}^2+(\hat{d}^{\dagger})^2-2\hat{d}^{\dagger}\hat{d}-1\right)|r\rangle.
\end{eqnarray}
Furthermore we have
\begin{eqnarray}
\langle r|\hat{c}^2|r\rangle&=&\frac{1}{2}\langle r|\left(\hat{a}^2+\hat{b}^2-2\hat{a}\hat{b}\right)|r\rangle, \nonumber \\
\langle r|\hat{c}^{\dagger}\hat{c}|r\rangle&=&\frac{1}{2}\langle r|\left(\hat{a}^{\dagger}\hat{a}+\hat{b}^{\dagger}\hat{b}-\hat{a}^{\dagger}\hat{b}-\hat{b}^{\dagger}\hat{a}\right)|r\rangle, \nonumber \\
\end{eqnarray}
and
\begin{eqnarray}
\langle r|\hat{d}^2|r\rangle&=&\frac{1}{2}\langle r|\left(\hat{a}^2+\hat{b}^2+2\hat{a}\hat{b}\right)|r\rangle, \nonumber \\
\langle r|\hat{d}^{\dagger}\hat{d}|r\rangle&=&\frac{1}{2}\langle r|\left(\hat{a}^{\dagger}\hat{a}+\hat{b}^{\dagger}\hat{b}+\hat{a}^{\dagger}\hat{b}+\hat{b}^{\dagger}\hat{a}\right)|r\rangle. \nonumber \\
\end{eqnarray}
In general we have the result $\langle r|F(\hat{a},\hat{a}^{\dagger};\hat{b},\hat{b}^{\dagger})|r\rangle=\langle vac|F(\hat{\alpha},\hat{\alpha}^{\dagger};\hat{\beta},\hat{\beta}^{\dagger})|vac\rangle$. We got before, equations (\ref{EX1},\ref{EX2}),
\begin{eqnarray}
\langle r|\hat{a}^2|r\rangle&=&\langle vac|\hat{\alpha}^2|vac\rangle=0, \nonumber \\
\langle r|\hat{a}^{\dagger}\hat{a}|r\rangle&=&\langle vac|\hat{\alpha}^{\dagger}\hat{\alpha}|vac\rangle=\sinh^2r,
\end{eqnarray}
and
\begin{eqnarray}
\langle r|\hat{b}^2|r\rangle&=&\langle vac|\hat{\beta}^2|vac\rangle=0, \nonumber \\
\langle r|\hat{b}^{\dagger}\hat{b}|r\rangle&=&\langle vac|\hat{\beta}^{\dagger}\hat{\beta}|vac\rangle=\sinh^2r.
\end{eqnarray}
We need also
\begin{equation}
\langle r|\hat{a}^{\dagger}\hat{b}|r\rangle=\langle vac|\hat{\alpha}^{\dagger}\hat{\beta}|vac\rangle,\ \langle r|\hat{a}\hat{b}|r\rangle=\langle vac|\hat{\alpha}\hat{\beta}|vac\rangle,
\end{equation}
where we have
\begin{eqnarray}
\hat{\alpha}^{\dagger}\hat{\beta}&=&\cosh r\sinh r((\hat{a}^{\dagger})^2+\hat{b}^2) \nonumber \\
&+&(\cosh^2r+\sinh^2r)\ \hat{a}^{\dagger}\hat{b}, \nonumber \\
\hat{\alpha}\hat{\beta}&=&\cosh r\sinh r(\hat{a}\hat{a}^{\dagger}+\hat{b}^{\dagger}\hat{b}) \nonumber \\
&+&\cosh^2r\ \hat{a}\hat{b}+\sinh^2r\ \hat{a}^{\dagger}\hat{b}^{\dagger}. \nonumber \\
\end{eqnarray}
Using (\ref{Ex3},\ref{Ex4},\ref{Ex5}), we obtain
\begin{equation}
\langle r|\hat{a}^{\dagger}\hat{b}|r\rangle=0,\ \langle r|\hat{a}\hat{b}|r\rangle=\cosh r\sinh r.
\end{equation}
We get
\begin{equation}
\langle r|\hat{c}^2|r\rangle=-\cosh r\sinh r,\ \langle r|\hat{c}^{\dagger}\hat{c}|r\rangle=\sinh^2r,
\end{equation}
and
\begin{equation}
\langle r|\hat{d}^2|r\rangle=\cosh r\sinh r,\ \langle r|\hat{d}^{\dagger}\hat{d}|r\rangle=\sinh^2r.
\end{equation}
Finally we reach
\begin{eqnarray}
\langle\hat{X}_c^2\rangle&=&\frac{1}{2}-\cosh r\sinh r+\sinh^2r, \nonumber \\
\langle\hat{Y}_c^2\rangle&=&\frac{1}{2}+\cosh r\sinh r+\sinh^2r, \nonumber \\
\langle\hat{X}_d^2\rangle&=&\frac{1}{2}+\cosh r\sinh r+\sinh^2r, \nonumber \\
\langle\hat{Y}_d^2\rangle&=&\frac{1}{2}-\cosh r\sinh r+\sinh^2r.
\end{eqnarray}


\begin{thebibliography}{52}
\expandafter\ifx\csname natexlab\endcsname\relax\def\natexlab#1{#1}\fi
\expandafter\ifx\csname bibnamefont\endcsname\relax
  \def\bibnamefont#1{#1}\fi
\expandafter\ifx\csname bibfnamefont\endcsname\relax
  \def\bibfnamefont#1{#1}\fi
\expandafter\ifx\csname citenamefont\endcsname\relax
  \def\citenamefont#1{#1}\fi
\expandafter\ifx\csname url\endcsname\relax
  \def\url#1{\texttt{#1}}\fi
\expandafter\ifx\csname urlprefix\endcsname\relax\def\urlprefix{URL }\fi
\providecommand{\bibinfo}[2]{#2}
\providecommand{\eprint}[2][]{\url{#2}}

\bibitem[{\citenamefont{Aspelmeyer et~al.}(2014)\citenamefont{Aspelmeyer,
  Kippenberg, and Marquardt}}]{Aspelmeyer2014}
\bibinfo{author}{\bibfnamefont{M.}~\bibnamefont{Aspelmeyer}},
  \bibinfo{author}{\bibfnamefont{T.~J.} \bibnamefont{Kippenberg}},
  \bibnamefont{and}
  \bibinfo{author}{\bibfnamefont{F.}~\bibnamefont{Marquardt}},
  \bibinfo{journal}{Rev. Mod. Phys.} \textbf{\bibinfo{volume}{86}},
  \bibinfo{pages}{1391} (\bibinfo{year}{2014}).

\bibitem[{\citenamefont{Clerk et~al.}(2010)\citenamefont{Clerk, Devoret,
  Girvin, Marquardt, and Schoelkopf}}]{Clerk2010}
\bibinfo{author}{\bibfnamefont{A.~A.} \bibnamefont{Clerk}},
  \bibinfo{author}{\bibfnamefont{M.~H.} \bibnamefont{Devoret}},
  \bibinfo{author}{\bibfnamefont{S.~M.} \bibnamefont{Girvin}},
  \bibinfo{author}{\bibfnamefont{F.}~\bibnamefont{Marquardt}},
  \bibnamefont{and} \bibinfo{author}{\bibfnamefont{R.~J.}
  \bibnamefont{Schoelkopf}}, \bibinfo{journal}{Rev. Mod. Phys.}
  \textbf{\bibinfo{volume}{82}}, \bibinfo{pages}{1155} (\bibinfo{year}{2010}).

\bibitem[{\citenamefont{Safavi-Naeini et~al.}(2019)\citenamefont{Safavi-Naeini,
  Van-Thourhout, Baets, and Van-Laer}}]{Safavi2019}
\bibinfo{author}{\bibfnamefont{A.~H.} \bibnamefont{Safavi-Naeini}},
  \bibinfo{author}{\bibfnamefont{D.}~\bibnamefont{Van-Thourhout}},
  \bibinfo{author}{\bibfnamefont{R.}~\bibnamefont{Baets}}, \bibnamefont{and}
  \bibinfo{author}{\bibfnamefont{R.}~\bibnamefont{Van-Laer}},
  \bibinfo{journal}{Optica} \textbf{\bibinfo{volume}{6}}, \bibinfo{pages}{213}
  (\bibinfo{year}{2019}).

\bibitem[{\citenamefont{Mancini and Tombesi}(1994)}]{Mancini1994}
\bibinfo{author}{\bibfnamefont{S.}~\bibnamefont{Mancini}} \bibnamefont{and}
  \bibinfo{author}{\bibfnamefont{P.}~\bibnamefont{Tombesi}},
  \bibinfo{journal}{Phys. Rev. A} \textbf{\bibinfo{volume}{49}},
  \bibinfo{pages}{4055} (\bibinfo{year}{1994}).

\bibitem[{\citenamefont{Hofer et~al.}(2011)\citenamefont{Hofer, Wieczorek,
  Aspelmeyer, and Hammerer}}]{Hofer2011}
\bibinfo{author}{\bibfnamefont{S.~G.} \bibnamefont{Hofer}},
  \bibinfo{author}{\bibfnamefont{W.}~\bibnamefont{Wieczorek}},
  \bibinfo{author}{\bibfnamefont{M.}~\bibnamefont{Aspelmeyer}},
  \bibnamefont{and} \bibinfo{author}{\bibfnamefont{K.}~\bibnamefont{Hammerer}},
  \bibinfo{journal}{Phys. Rev. A} \textbf{\bibinfo{volume}{84}},
  \bibinfo{pages}{052327} (\bibinfo{year}{2011}).

\bibitem[{\citenamefont{Riedinger et~al.}(2016)\citenamefont{Riedinger, Hong,
  Norte, Slater, Shang, Krause, Anant, Aspelmeyer, and
  Gröblacher}}]{Riedinger2016}
\bibinfo{author}{\bibfnamefont{R.}~\bibnamefont{Riedinger}},
  \bibinfo{author}{\bibfnamefont{S.}~\bibnamefont{Hong}},
  \bibinfo{author}{\bibfnamefont{R.~A.} \bibnamefont{Norte}},
  \bibinfo{author}{\bibfnamefont{J.~A.} \bibnamefont{Slater}},
  \bibinfo{author}{\bibfnamefont{J.}~\bibnamefont{Shang}},
  \bibinfo{author}{\bibfnamefont{A.~G.} \bibnamefont{Krause}},
  \bibinfo{author}{\bibfnamefont{V.}~\bibnamefont{Anant}},
  \bibinfo{author}{\bibfnamefont{M.}~\bibnamefont{Aspelmeyer}},
  \bibnamefont{and}
  \bibinfo{author}{\bibfnamefont{S.}~\bibnamefont{Gröblacher}},
  \bibinfo{journal}{Nature} \textbf{\bibinfo{volume}{530}},
  \bibinfo{pages}{313} (\bibinfo{year}{2016}).

\bibitem[{\citenamefont{Mari and Eisert}(2009)}]{Mari2009}
\bibinfo{author}{\bibfnamefont{A.}~\bibnamefont{Mari}} \bibnamefont{and}
  \bibinfo{author}{\bibfnamefont{J.}~\bibnamefont{Eisert}},
  \bibinfo{journal}{Phys. Rev. Lett.} \textbf{\bibinfo{volume}{103}},
  \bibinfo{pages}{213603} (\bibinfo{year}{2009}).

\bibitem[{\citenamefont{Liao and Law}(2011)}]{Liao2011}
\bibinfo{author}{\bibfnamefont{J.-Q.} \bibnamefont{Liao}} \bibnamefont{and}
  \bibinfo{author}{\bibfnamefont{C.~K.} \bibnamefont{Law}},
  \bibinfo{journal}{Phys. Rev. A} \textbf{\bibinfo{volume}{83}},
  \bibinfo{pages}{033820} (\bibinfo{year}{2011}).

\bibitem[{\citenamefont{Farace and Giovannetti}(2012)}]{Farace2012}
\bibinfo{author}{\bibfnamefont{A.}~\bibnamefont{Farace}} \bibnamefont{and}
  \bibinfo{author}{\bibfnamefont{V.}~\bibnamefont{Giovannetti}},
  \bibinfo{journal}{Phys. Rev. A} \textbf{\bibinfo{volume}{86}},
  \bibinfo{pages}{013820} (\bibinfo{year}{2012}).

\bibitem[{\citenamefont{Szorkovszky et~al.}(2011)\citenamefont{Szorkovszky,
  Doherty, Harris, and Bowen}}]{Szorkovszky2011}
\bibinfo{author}{\bibfnamefont{A.}~\bibnamefont{Szorkovszky}},
  \bibinfo{author}{\bibfnamefont{A.~C.} \bibnamefont{Doherty}},
  \bibinfo{author}{\bibfnamefont{G.~I.} \bibnamefont{Harris}},
  \bibnamefont{and} \bibinfo{author}{\bibfnamefont{W.~P.} \bibnamefont{Bowen}},
  \bibinfo{journal}{Phys. Rev. Lett.} \textbf{\bibinfo{volume}{107}},
  \bibinfo{pages}{213603} (\bibinfo{year}{2011}).

\bibitem[{\citenamefont{Schmidt et~al.}(2012)\citenamefont{Schmidt, Ludwig, and
  Marquardt}}]{Schmidt2012}
\bibinfo{author}{\bibfnamefont{M.}~\bibnamefont{Schmidt}},
  \bibinfo{author}{\bibfnamefont{M.}~\bibnamefont{Ludwig}}, \bibnamefont{and}
  \bibinfo{author}{\bibfnamefont{F.}~\bibnamefont{Marquardt}},
  \bibinfo{journal}{New Journal of Physics} \textbf{\bibinfo{volume}{14}},
  \bibinfo{pages}{125005} (\bibinfo{year}{2012}).

\bibitem[{\citenamefont{Paternostro et~al.}(2007)\citenamefont{Paternostro,
  Vitali, Gigan, Kim, Brukner, Eisert, and Aspelmeyer}}]{Paternostro2007}
\bibinfo{author}{\bibfnamefont{M.}~\bibnamefont{Paternostro}},
  \bibinfo{author}{\bibfnamefont{D.}~\bibnamefont{Vitali}},
  \bibinfo{author}{\bibfnamefont{S.}~\bibnamefont{Gigan}},
  \bibinfo{author}{\bibfnamefont{M.~S.} \bibnamefont{Kim}},
  \bibinfo{author}{\bibfnamefont{C.}~\bibnamefont{Brukner}},
  \bibinfo{author}{\bibfnamefont{J.}~\bibnamefont{Eisert}}, \bibnamefont{and}
  \bibinfo{author}{\bibfnamefont{M.}~\bibnamefont{Aspelmeyer}},
  \bibinfo{journal}{Phys. Rev. Lett.} \textbf{\bibinfo{volume}{99}},
  \bibinfo{pages}{250401} (\bibinfo{year}{2007}).

\bibitem[{\citenamefont{Vitali et~al.}(2007)\citenamefont{Vitali, Gigan,
  Ferreira, B\"ohm, Tombesi, Guerreiro, Vedral, Zeilinger, and
  Aspelmeyer}}]{Vitali2007}
\bibinfo{author}{\bibfnamefont{D.}~\bibnamefont{Vitali}},
  \bibinfo{author}{\bibfnamefont{S.}~\bibnamefont{Gigan}},
  \bibinfo{author}{\bibfnamefont{A.}~\bibnamefont{Ferreira}},
  \bibinfo{author}{\bibfnamefont{H.~R.} \bibnamefont{B\"ohm}},
  \bibinfo{author}{\bibfnamefont{P.}~\bibnamefont{Tombesi}},
  \bibinfo{author}{\bibfnamefont{A.}~\bibnamefont{Guerreiro}},
  \bibinfo{author}{\bibfnamefont{V.}~\bibnamefont{Vedral}},
  \bibinfo{author}{\bibfnamefont{A.}~\bibnamefont{Zeilinger}},
  \bibnamefont{and}
  \bibinfo{author}{\bibfnamefont{M.}~\bibnamefont{Aspelmeyer}},
  \bibinfo{journal}{Phys. Rev. Lett.} \textbf{\bibinfo{volume}{98}},
  \bibinfo{pages}{030405} (\bibinfo{year}{2007}).

\bibitem[{\citenamefont{Palomaki et~al.}(2013)\citenamefont{Palomaki, Teufel,
  Simmonds, and Lehnert}}]{Palomaki2013}
\bibinfo{author}{\bibfnamefont{T.~A.} \bibnamefont{Palomaki}},
  \bibinfo{author}{\bibfnamefont{J.~D.} \bibnamefont{Teufel}},
  \bibinfo{author}{\bibfnamefont{R.~W.} \bibnamefont{Simmonds}},
  \bibnamefont{and} \bibinfo{author}{\bibfnamefont{K.~W.}
  \bibnamefont{Lehnert}}, \bibinfo{journal}{Science}
  \textbf{\bibinfo{volume}{342}}, \bibinfo{pages}{710} (\bibinfo{year}{2013}).

\bibitem[{\citenamefont{Yuen}(1976)}]{Yuen1976}
\bibinfo{author}{\bibfnamefont{H.~P.} \bibnamefont{Yuen}},
  \bibinfo{journal}{Phys. Rev. A} \textbf{\bibinfo{volume}{13}},
  \bibinfo{pages}{2226} (\bibinfo{year}{1976}).

\bibitem[{\citenamefont{Safavi-Naeini et~al.}(2013)\citenamefont{Safavi-Naeini,
  Gröblacher, Hill, Chan, Aspelmeyer, and Painter}}]{Safavi2013}
\bibinfo{author}{\bibfnamefont{A.~H.} \bibnamefont{Safavi-Naeini}},
  \bibinfo{author}{\bibfnamefont{S.}~\bibnamefont{Gröblacher}},
  \bibinfo{author}{\bibfnamefont{J.~T.} \bibnamefont{Hill}},
  \bibinfo{author}{\bibfnamefont{J.}~\bibnamefont{Chan}},
  \bibinfo{author}{\bibfnamefont{M.}~\bibnamefont{Aspelmeyer}},
  \bibnamefont{and} \bibinfo{author}{\bibfnamefont{O.}~\bibnamefont{Painter}},
  \bibinfo{journal}{Nature} \textbf{\bibinfo{volume}{500}},
  \bibinfo{pages}{185} (\bibinfo{year}{2013}).

\bibitem[{\citenamefont{Wollman et~al.}(2015)\citenamefont{Wollman, Lei,
  Weinstein, Suh, Kronwald, Marquardt, Clerk, and Schwab}}]{Wollman2015}
\bibinfo{author}{\bibfnamefont{E.~E.} \bibnamefont{Wollman}},
  \bibinfo{author}{\bibfnamefont{C.~C.} \bibnamefont{Lei}},
  \bibinfo{author}{\bibfnamefont{A.~J.} \bibnamefont{Weinstein}},
  \bibinfo{author}{\bibfnamefont{J.}~\bibnamefont{Suh}},
  \bibinfo{author}{\bibfnamefont{A.}~\bibnamefont{Kronwald}},
  \bibinfo{author}{\bibfnamefont{F.}~\bibnamefont{Marquardt}},
  \bibinfo{author}{\bibfnamefont{A.~A.} \bibnamefont{Clerk}}, \bibnamefont{and}
  \bibinfo{author}{\bibfnamefont{K.~C.} \bibnamefont{Schwab}},
  \bibinfo{journal}{Science} \textbf{\bibinfo{volume}{349}},
  \bibinfo{pages}{952} (\bibinfo{year}{2015}).

\bibitem[{\citenamefont{Chu et~al.}(2018)\citenamefont{Chu, Kharel, Yoon,
  Frunzio, Rakich, and Schoelkopf}}]{Chu2018}
\bibinfo{author}{\bibfnamefont{Y.}~\bibnamefont{Chu}},
  \bibinfo{author}{\bibfnamefont{P.}~\bibnamefont{Kharel}},
  \bibinfo{author}{\bibfnamefont{T.}~\bibnamefont{Yoon}},
  \bibinfo{author}{\bibfnamefont{L.}~\bibnamefont{Frunzio}},
  \bibinfo{author}{\bibfnamefont{P.~T.} \bibnamefont{Rakich}},
  \bibnamefont{and} \bibinfo{author}{\bibfnamefont{R.~J.}
  \bibnamefont{Schoelkopf}}, \bibinfo{journal}{Nature}
  \textbf{\bibinfo{volume}{563}}, \bibinfo{pages}{666} (\bibinfo{year}{2018}).

\bibitem[{\citenamefont{Duan et~al.}(2001)\citenamefont{Duan, Lukin, Cirac, and
  Zoller}}]{Duan2001}
\bibinfo{author}{\bibfnamefont{L.~M.} \bibnamefont{Duan}},
  \bibinfo{author}{\bibfnamefont{M.~D.} \bibnamefont{Lukin}},
  \bibinfo{author}{\bibfnamefont{J.~I.} \bibnamefont{Cirac}}, \bibnamefont{and}
  \bibinfo{author}{\bibfnamefont{P.}~\bibnamefont{Zoller}},
  \bibinfo{journal}{Nature} \textbf{\bibinfo{volume}{414}},
  \bibinfo{pages}{413} (\bibinfo{year}{2001}).

\bibitem[{\citenamefont{Stannigel et~al.}(2010)\citenamefont{Stannigel, Rabl,
  Sorensen, Zoller, and Lukin}}]{Stannigel2010}
\bibinfo{author}{\bibfnamefont{K.}~\bibnamefont{Stannigel}},
  \bibinfo{author}{\bibfnamefont{P.}~\bibnamefont{Rabl}},
  \bibinfo{author}{\bibfnamefont{A.~S.} \bibnamefont{Sorensen}},
  \bibinfo{author}{\bibfnamefont{P.}~\bibnamefont{Zoller}}, \bibnamefont{and}
  \bibinfo{author}{\bibfnamefont{M.~D.} \bibnamefont{Lukin}},
  \bibinfo{journal}{Phys. Rev. Lett.} \textbf{\bibinfo{volume}{105}},
  \bibinfo{pages}{220501} (\bibinfo{year}{2010}).

\bibitem[{\citenamefont{Pirkkalainen et~al.}(2015)\citenamefont{Pirkkalainen,
  Cho, Massel, Tuorila, Heikkilä, Hakonen, and
  Sillanpää}}]{Pirkkalainen2015}
\bibinfo{author}{\bibfnamefont{L.~M.} \bibnamefont{Pirkkalainen}},
  \bibinfo{author}{\bibfnamefont{S.~U.} \bibnamefont{Cho}},
  \bibinfo{author}{\bibfnamefont{F.}~\bibnamefont{Massel}},
  \bibinfo{author}{\bibfnamefont{J.}~\bibnamefont{Tuorila}},
  \bibinfo{author}{\bibfnamefont{T.~T.} \bibnamefont{Heikkilä}},
  \bibinfo{author}{\bibfnamefont{P.~J.} \bibnamefont{Hakonen}},
  \bibnamefont{and} \bibinfo{author}{\bibfnamefont{M.~A.}
  \bibnamefont{Sillanpää}}, \bibinfo{journal}{Nature Communications}
  \textbf{\bibinfo{volume}{6}}, \bibinfo{pages}{6981} (\bibinfo{year}{2015}).

\bibitem[{\citenamefont{Lemonde et~al.}(2016)\citenamefont{Lemonde, Didier, and
  Clerk}}]{Lemonde2016}
\bibinfo{author}{\bibfnamefont{M.-A.} \bibnamefont{Lemonde}},
  \bibinfo{author}{\bibfnamefont{N.}~\bibnamefont{Didier}}, \bibnamefont{and}
  \bibinfo{author}{\bibfnamefont{A.~A.} \bibnamefont{Clerk}},
  \bibinfo{journal}{Nature Communications} \textbf{\bibinfo{volume}{7}},
  \bibinfo{pages}{11338} (\bibinfo{year}{2016}).

\bibitem[{\citenamefont{Mirza}(2016)}]{Mirza2016}
\bibinfo{author}{\bibfnamefont{I.~M.} \bibnamefont{Mirza}},
  \bibinfo{journal}{Opt. Lett.} \textbf{\bibinfo{volume}{41}},
  \bibinfo{pages}{2422} (\bibinfo{year}{2016}).

\bibitem[{\citenamefont{Riedinger et~al.}(2018)\citenamefont{Riedinger,
  Wallucks, Marinkovic, Löschnauer, Aspelmeyer, Hong, and
  Gröblacher}}]{Riedinger2018}
\bibinfo{author}{\bibfnamefont{R.}~\bibnamefont{Riedinger}},
  \bibinfo{author}{\bibfnamefont{A.}~\bibnamefont{Wallucks}},
  \bibinfo{author}{\bibfnamefont{I.}~\bibnamefont{Marinkovic}},
  \bibinfo{author}{\bibfnamefont{C.}~\bibnamefont{Löschnauer}},
  \bibinfo{author}{\bibfnamefont{M.}~\bibnamefont{Aspelmeyer}},
  \bibinfo{author}{\bibfnamefont{S.}~\bibnamefont{Hong}}, \bibnamefont{and}
  \bibinfo{author}{\bibfnamefont{S.}~\bibnamefont{Gröblacher}},
  \bibinfo{journal}{Nature} \textbf{\bibinfo{volume}{556}},
  \bibinfo{pages}{473} (\bibinfo{year}{2018}).

\bibitem[{\citenamefont{Boyd}(2008)}]{Boyd2008}
\bibinfo{author}{\bibfnamefont{R.~W.} \bibnamefont{Boyd}},
  \emph{\bibinfo{title}{Nonlinear Optics}} (\bibinfo{publisher}{Elsevier},
  \bibinfo{address}{Amsterdam}, \bibinfo{year}{2008}), \bibinfo{edition}{3rd}
  ed.

\bibitem[{\citenamefont{Agrawal}(2013)}]{Agrawal2013}
\bibinfo{author}{\bibfnamefont{G.~P.} \bibnamefont{Agrawal}},
  \emph{\bibinfo{title}{Nonlinear Fiber Optics}}
  (\bibinfo{publisher}{Elsevier}, \bibinfo{address}{Amsterdam},
  \bibinfo{year}{2013}), \bibinfo{edition}{5th} ed.

\bibitem[{\citenamefont{Rakich et~al.}(2012)\citenamefont{Rakich, Reinke,
  Camacho, Davids, and Wang}}]{Rakish2012}
\bibinfo{author}{\bibfnamefont{P.~T.} \bibnamefont{Rakich}},
  \bibinfo{author}{\bibfnamefont{C.}~\bibnamefont{Reinke}},
  \bibinfo{author}{\bibfnamefont{R.}~\bibnamefont{Camacho}},
  \bibinfo{author}{\bibfnamefont{P.}~\bibnamefont{Davids}}, \bibnamefont{and}
  \bibinfo{author}{\bibfnamefont{Z.}~\bibnamefont{Wang}},
  \bibinfo{journal}{Phys. Rev. X} \textbf{\bibinfo{volume}{2}},
  \bibinfo{pages}{011008} (\bibinfo{year}{2012}).

\bibitem[{\citenamefont{Van-Laer
  et~al.}(2015{\natexlab{a}})\citenamefont{Van-Laer, Kuyken, Van-Thourhout, and
  Baets}}]{VanLaer2015a}
\bibinfo{author}{\bibfnamefont{R.}~\bibnamefont{Van-Laer}},
  \bibinfo{author}{\bibfnamefont{B.}~\bibnamefont{Kuyken}},
  \bibinfo{author}{\bibfnamefont{D.}~\bibnamefont{Van-Thourhout}},
  \bibnamefont{and} \bibinfo{author}{\bibfnamefont{R.}~\bibnamefont{Baets}},
  \bibinfo{journal}{Nature Photonics} \textbf{\bibinfo{volume}{9}},
  \bibinfo{pages}{199} (\bibinfo{year}{2015}{\natexlab{a}}).

\bibitem[{\citenamefont{Zoubi and Hammerer}(2016)}]{Zoubi2016}
\bibinfo{author}{\bibfnamefont{H.}~\bibnamefont{Zoubi}} \bibnamefont{and}
  \bibinfo{author}{\bibfnamefont{K.}~\bibnamefont{Hammerer}},
  \bibinfo{journal}{Phys. Rev. A} \textbf{\bibinfo{volume}{94}},
  \bibinfo{pages}{053827} (\bibinfo{year}{2016}).

\bibitem[{\citenamefont{Thevenaz}(2008)}]{Thevenaz2008}
\bibinfo{author}{\bibfnamefont{L.}~\bibnamefont{Thevenaz}},
  \bibinfo{journal}{Nature Photonics} \textbf{\bibinfo{volume}{2}},
  \bibinfo{pages}{474} (\bibinfo{year}{2008}).

\bibitem[{\citenamefont{Bahl et~al.}(2012)\citenamefont{Bahl, Tomes, Marquardt,
  and Carmon}}]{Bahl2012}
\bibinfo{author}{\bibfnamefont{G.}~\bibnamefont{Bahl}},
  \bibinfo{author}{\bibfnamefont{M.}~\bibnamefont{Tomes}},
  \bibinfo{author}{\bibfnamefont{F.}~\bibnamefont{Marquardt}},
  \bibnamefont{and} \bibinfo{author}{\bibfnamefont{T.}~\bibnamefont{Carmon}},
  \bibinfo{journal}{Nature Physics} \textbf{\bibinfo{volume}{8}},
  \bibinfo{pages}{203} (\bibinfo{year}{2012}).

\bibitem[{\citenamefont{Agarwal and Jha}(2013)}]{Agarwal2013a}
\bibinfo{author}{\bibfnamefont{G.~S.} \bibnamefont{Agarwal}} \bibnamefont{and}
  \bibinfo{author}{\bibfnamefont{S.~S.} \bibnamefont{Jha}},
  \bibinfo{journal}{Phys. Rev. A} \textbf{\bibinfo{volume}{88}},
  \bibinfo{pages}{013815} (\bibinfo{year}{2013}).

\bibitem[{\citenamefont{Beugnot et~al.}(2014)\citenamefont{Beugnot, Lebrun,
  Pauliat, Maillotte, Laude, and Sylvestre}}]{Beugnot2014}
\bibinfo{author}{\bibfnamefont{J.-C.} \bibnamefont{Beugnot}},
  \bibinfo{author}{\bibfnamefont{S.}~\bibnamefont{Lebrun}},
  \bibinfo{author}{\bibfnamefont{G.}~\bibnamefont{Pauliat}},
  \bibinfo{author}{\bibfnamefont{H.}~\bibnamefont{Maillotte}},
  \bibinfo{author}{\bibfnamefont{V.}~\bibnamefont{Laude}}, \bibnamefont{and}
  \bibinfo{author}{\bibfnamefont{T.}~\bibnamefont{Sylvestre}},
  \bibinfo{journal}{Nature Communications} \textbf{\bibinfo{volume}{5}},
  \bibinfo{pages}{5242} (\bibinfo{year}{2014}).

\bibitem[{\citenamefont{Merklein et~al.}(2017)\citenamefont{Merklein, Stiller,
  Vu, Madden, and Eggleton}}]{Merklein2016}
\bibinfo{author}{\bibfnamefont{M.}~\bibnamefont{Merklein}},
  \bibinfo{author}{\bibfnamefont{B.}~\bibnamefont{Stiller}},
  \bibinfo{author}{\bibfnamefont{K.}~\bibnamefont{Vu}},
  \bibinfo{author}{\bibfnamefont{S.~J.} \bibnamefont{Madden}},
  \bibnamefont{and} \bibinfo{author}{\bibfnamefont{B.~J.}
  \bibnamefont{Eggleton}}, \bibinfo{journal}{nature Communications}
  \textbf{\bibinfo{volume}{8}}, \bibinfo{pages}{574} (\bibinfo{year}{2017}).

\bibitem[{\citenamefont{Zoubi and Hammerer}(2017)}]{Zoubi2017}
\bibinfo{author}{\bibfnamefont{H.}~\bibnamefont{Zoubi}} \bibnamefont{and}
  \bibinfo{author}{\bibfnamefont{K.}~\bibnamefont{Hammerer}},
  \bibinfo{journal}{Physical Review Letters} \textbf{\bibinfo{volume}{119}},
  \bibinfo{pages}{123602} (\bibinfo{year}{2017}).

\bibitem[{\citenamefont{Zoubi}(2018)}]{Zoubi2018}
\bibinfo{author}{\bibfnamefont{H.}~\bibnamefont{Zoubi}},
  \bibinfo{journal}{Journal of Optics} \textbf{\bibinfo{volume}{20}},
  \bibinfo{pages}{095001} (\bibinfo{year}{2018}).

\bibitem[{\citenamefont{Eggleton et~al.}(2013)\citenamefont{Eggleton, Poulton,
  and Pant}}]{Eggleton2013}
\bibinfo{author}{\bibfnamefont{B.~J.} \bibnamefont{Eggleton}},
  \bibinfo{author}{\bibfnamefont{C.~G.} \bibnamefont{Poulton}},
  \bibnamefont{and} \bibinfo{author}{\bibfnamefont{R.}~\bibnamefont{Pant}},
  \bibinfo{journal}{Adv. Opt. Photon.} \textbf{\bibinfo{volume}{5}},
  \bibinfo{pages}{536} (\bibinfo{year}{2013}).

\bibitem[{\citenamefont{Safavi-Naeini et~al.}(2011)\citenamefont{Safavi-Naeini,
  Alegre, Chan, Eichenfield, Winger, Lin, Hill, Chang, and
  Painter}}]{Safavi-Naeini2011}
\bibinfo{author}{\bibfnamefont{A.~H.} \bibnamefont{Safavi-Naeini}},
  \bibinfo{author}{\bibfnamefont{T.~P.~M.} \bibnamefont{Alegre}},
  \bibinfo{author}{\bibfnamefont{J.}~\bibnamefont{Chan}},
  \bibinfo{author}{\bibfnamefont{M.}~\bibnamefont{Eichenfield}},
  \bibinfo{author}{\bibfnamefont{M.}~\bibnamefont{Winger}},
  \bibinfo{author}{\bibfnamefont{Q.}~\bibnamefont{Lin}},
  \bibinfo{author}{\bibfnamefont{J.~T.} \bibnamefont{Hill}},
  \bibinfo{author}{\bibfnamefont{D.~E.} \bibnamefont{Chang}}, \bibnamefont{and}
  \bibinfo{author}{\bibfnamefont{O.}~\bibnamefont{Painter}},
  \bibinfo{journal}{Nature} \textbf{\bibinfo{volume}{472}}, \bibinfo{pages}{69}
  (\bibinfo{year}{2011}).

\bibitem[{\citenamefont{Weis et~al.}(2010)\citenamefont{Weis, Rivi{\`e}re,
  Del{\'e}glise, Gavartin, Arcizet, Schliesser, and Kippenberg}}]{Weis2010}
\bibinfo{author}{\bibfnamefont{S.}~\bibnamefont{Weis}},
  \bibinfo{author}{\bibfnamefont{R.}~\bibnamefont{Rivi{\`e}re}},
  \bibinfo{author}{\bibfnamefont{S.}~\bibnamefont{Del{\'e}glise}},
  \bibinfo{author}{\bibfnamefont{E.}~\bibnamefont{Gavartin}},
  \bibinfo{author}{\bibfnamefont{O.}~\bibnamefont{Arcizet}},
  \bibinfo{author}{\bibfnamefont{A.}~\bibnamefont{Schliesser}},
  \bibnamefont{and} \bibinfo{author}{\bibfnamefont{T.~J.}
  \bibnamefont{Kippenberg}}, \bibinfo{journal}{Science}
  \textbf{\bibinfo{volume}{330}}, \bibinfo{pages}{1520} (\bibinfo{year}{2010}).

\bibitem[{\citenamefont{Kim et~al.}(2015)\citenamefont{Kim, Kuzyk, Han, Wang,
  and Bahl}}]{Kim2015}
\bibinfo{author}{\bibfnamefont{J.}~\bibnamefont{Kim}},
  \bibinfo{author}{\bibfnamefont{M.~C.} \bibnamefont{Kuzyk}},
  \bibinfo{author}{\bibfnamefont{K.}~\bibnamefont{Han}},
  \bibinfo{author}{\bibfnamefont{H.}~\bibnamefont{Wang}}, \bibnamefont{and}
  \bibinfo{author}{\bibfnamefont{G.}~\bibnamefont{Bahl}},
  \bibinfo{journal}{Nature Physics} \textbf{\bibinfo{volume}{11}},
  \bibinfo{pages}{275} (\bibinfo{year}{2015}).

\bibitem[{\citenamefont{Pant et~al.}(2011)\citenamefont{Pant, Poulton, Choi,
  Mcfarlane, Hile, Li, Thevenaz, Luther-Davies, Madden, and
  Eggleton}}]{Pant2011}
\bibinfo{author}{\bibfnamefont{R.}~\bibnamefont{Pant}},
  \bibinfo{author}{\bibfnamefont{C.~G.} \bibnamefont{Poulton}},
  \bibinfo{author}{\bibfnamefont{D.-Y.} \bibnamefont{Choi}},
  \bibinfo{author}{\bibfnamefont{H.}~\bibnamefont{Mcfarlane}},
  \bibinfo{author}{\bibfnamefont{S.}~\bibnamefont{Hile}},
  \bibinfo{author}{\bibfnamefont{E.}~\bibnamefont{Li}},
  \bibinfo{author}{\bibfnamefont{L.}~\bibnamefont{Thevenaz}},
  \bibinfo{author}{\bibfnamefont{B.}~\bibnamefont{Luther-Davies}},
  \bibinfo{author}{\bibfnamefont{S.~J.} \bibnamefont{Madden}},
  \bibnamefont{and} \bibinfo{author}{\bibfnamefont{B.~J.}
  \bibnamefont{Eggleton}}, \bibinfo{journal}{Opt. Express}
  \textbf{\bibinfo{volume}{19}}, \bibinfo{pages}{8285} (\bibinfo{year}{2011}).

\bibitem[{\citenamefont{Shin et~al.}(2013)\citenamefont{Shin, Qiu, Jarecki,
  Cox, Olsson~III, Starbuck, Wang, and Rakich}}]{Shin2013}
\bibinfo{author}{\bibfnamefont{H.}~\bibnamefont{Shin}},
  \bibinfo{author}{\bibfnamefont{W.}~\bibnamefont{Qiu}},
  \bibinfo{author}{\bibfnamefont{R.}~\bibnamefont{Jarecki}},
  \bibinfo{author}{\bibfnamefont{J.~A.} \bibnamefont{Cox}},
  \bibinfo{author}{\bibfnamefont{R.~H.} \bibnamefont{Olsson~III}},
  \bibinfo{author}{\bibfnamefont{A.}~\bibnamefont{Starbuck}},
  \bibinfo{author}{\bibfnamefont{Z.}~\bibnamefont{Wang}}, \bibnamefont{and}
  \bibinfo{author}{\bibfnamefont{P.~T.} \bibnamefont{Rakich}},
  \bibinfo{journal}{Nature Communications} \textbf{\bibinfo{volume}{4}},
  \bibinfo{pages}{1944} (\bibinfo{year}{2013}).

\bibitem[{\citenamefont{Van-Laer
  et~al.}(2015{\natexlab{b}})\citenamefont{Van-Laer, Bazin, Kuyken, Baets, and
  Van-Thourhout}}]{VanLaer2015b}
\bibinfo{author}{\bibfnamefont{R.}~\bibnamefont{Van-Laer}},
  \bibinfo{author}{\bibfnamefont{A.}~\bibnamefont{Bazin}},
  \bibinfo{author}{\bibfnamefont{B.}~\bibnamefont{Kuyken}},
  \bibinfo{author}{\bibfnamefont{R.}~\bibnamefont{Baets}}, \bibnamefont{and}
  \bibinfo{author}{\bibfnamefont{D.}~\bibnamefont{Van-Thourhout}},
  \bibinfo{journal}{New Journal of Physics} \textbf{\bibinfo{volume}{17}},
  \bibinfo{pages}{115005} (\bibinfo{year}{2015}{\natexlab{b}}).

\bibitem[{\citenamefont{Kittlaus et~al.}(2015)\citenamefont{Kittlaus, Shin, and
  Rakich}}]{Kittlaus2015}
\bibinfo{author}{\bibfnamefont{E.~A.} \bibnamefont{Kittlaus}},
  \bibinfo{author}{\bibfnamefont{H.}~\bibnamefont{Shin}}, \bibnamefont{and}
  \bibinfo{author}{\bibfnamefont{P.~T.} \bibnamefont{Rakich}},
  \bibinfo{journal}{Nature Photonics} \textbf{\bibinfo{volume}{10}},
  \bibinfo{pages}{463} (\bibinfo{year}{2015}).

\bibitem[{\citenamefont{Loudon}(2000)}]{Loudon2000}
\bibinfo{author}{\bibfnamefont{R.}~\bibnamefont{Loudon}},
  \emph{\bibinfo{title}{The Quantum Theory of Light}}
  (\bibinfo{publisher}{Oxford}, \bibinfo{address}{UK}, \bibinfo{year}{2000}),
  \bibinfo{edition}{3rd} ed.

\bibitem[{\citenamefont{Walls and Milburn}(2008)}]{Walls2008}
\bibinfo{author}{\bibfnamefont{D.~F.} \bibnamefont{Walls}} \bibnamefont{and}
  \bibinfo{author}{\bibfnamefont{G.~J.} \bibnamefont{Milburn}},
  \emph{\bibinfo{title}{Quantum Optics}} (\bibinfo{publisher}{Springer-Verlag,
  Berlin}, \bibinfo{year}{2008}).

\bibitem[{\citenamefont{Gardiner and Zoller}(2010)}]{Gardiner2010}
\bibinfo{author}{\bibfnamefont{C.~W.} \bibnamefont{Gardiner}} \bibnamefont{and}
  \bibinfo{author}{\bibfnamefont{P.}~\bibnamefont{Zoller}},
  \emph{\bibinfo{title}{Quantum Noise}} (\bibinfo{publisher}{Springer-Verlag,
  Berlin}, \bibinfo{year}{2010}).

\bibitem[{\citenamefont{Vetsch et~al.}(2010)\citenamefont{Vetsch, Reitz, Sague,
  Schmidt, Dawkins, and Rauschenbeutel}}]{Vetsch2010}
\bibinfo{author}{\bibfnamefont{E.}~\bibnamefont{Vetsch}},
  \bibinfo{author}{\bibfnamefont{D.}~\bibnamefont{Reitz}},
  \bibinfo{author}{\bibfnamefont{G.}~\bibnamefont{Sague}},
  \bibinfo{author}{\bibfnamefont{R.}~\bibnamefont{Schmidt}},
  \bibinfo{author}{\bibfnamefont{S.~T.} \bibnamefont{Dawkins}},
  \bibnamefont{and}
  \bibinfo{author}{\bibfnamefont{A.}~\bibnamefont{Rauschenbeutel}},
  \bibinfo{journal}{Phys. Rev. Lett.} \textbf{\bibinfo{volume}{104}},
  \bibinfo{pages}{203603} (\bibinfo{year}{2010}).

\bibitem[{\citenamefont{Goban et~al.}(2012)\citenamefont{Goban, Choi, Alton,
  Ding, Lacroute, Pototschnig, Thiele, Stern, and Kimble}}]{Goban2012}
\bibinfo{author}{\bibfnamefont{A.}~\bibnamefont{Goban}},
  \bibinfo{author}{\bibfnamefont{K.~S.} \bibnamefont{Choi}},
  \bibinfo{author}{\bibfnamefont{D.~J.} \bibnamefont{Alton}},
  \bibinfo{author}{\bibfnamefont{D.}~\bibnamefont{Ding}},
  \bibinfo{author}{\bibfnamefont{C.}~\bibnamefont{Lacroute}},
  \bibinfo{author}{\bibfnamefont{M.}~\bibnamefont{Pototschnig}},
  \bibinfo{author}{\bibfnamefont{T.}~\bibnamefont{Thiele}},
  \bibinfo{author}{\bibfnamefont{N.~P.} \bibnamefont{Stern}}, \bibnamefont{and}
  \bibinfo{author}{\bibfnamefont{H.~J.} \bibnamefont{Kimble}},
  \bibinfo{journal}{Phys. Rev. Lett.} \textbf{\bibinfo{volume}{109}},
  \bibinfo{pages}{033603} (\bibinfo{year}{2012}).

\bibitem[{\citenamefont{Taillaert et~al.}(2006)\citenamefont{Taillaert,
  Van-Laere, Ayre, Bogaerts, Van-Thourhout, Bienstman, and
  Baets}}]{Taillaert2006}
\bibinfo{author}{\bibfnamefont{D.}~\bibnamefont{Taillaert}},
  \bibinfo{author}{\bibfnamefont{F.}~\bibnamefont{Van-Laere}},
  \bibinfo{author}{\bibfnamefont{M.}~\bibnamefont{Ayre}},
  \bibinfo{author}{\bibfnamefont{W.}~\bibnamefont{Bogaerts}},
  \bibinfo{author}{\bibfnamefont{D.}~\bibnamefont{Van-Thourhout}},
  \bibinfo{author}{\bibfnamefont{P.}~\bibnamefont{Bienstman}},
  \bibnamefont{and} \bibinfo{author}{\bibfnamefont{R.}~\bibnamefont{Baets}},
  \bibinfo{journal}{Japanese Journal of Applied Physics}
  \textbf{\bibinfo{volume}{45}}, \bibinfo{pages}{6071} (\bibinfo{year}{2006}).

\bibitem[{\citenamefont{Fetter and D}(1971)}]{Fetter1971}
\bibinfo{author}{\bibfnamefont{A.~L.} \bibnamefont{Fetter}} \bibnamefont{and}
  \bibinfo{author}{\bibfnamefont{W.~J.} \bibnamefont{D}},
  \emph{\bibinfo{title}{Quantum Theory of Many-Particle Systems}}
  (\bibinfo{publisher}{McGraw-Hill Book Company, New York},
  \bibinfo{year}{1971}).

\bibitem[{\citenamefont{Van-Laer et~al.}(2017)\citenamefont{Van-Laer,
  Sarabalis, Baets, Van-Thourhout, and Safavi-Naeini}}]{VanLaer2017}
\bibinfo{author}{\bibfnamefont{R.}~\bibnamefont{Van-Laer}},
  \bibinfo{author}{\bibfnamefont{C.~J.} \bibnamefont{Sarabalis}},
  \bibinfo{author}{\bibfnamefont{R.}~\bibnamefont{Baets}},
  \bibinfo{author}{\bibfnamefont{D.}~\bibnamefont{Van-Thourhout}},
  \bibnamefont{and} \bibinfo{author}{\bibfnamefont{A.~H.}
  \bibnamefont{Safavi-Naeini}}, \bibinfo{journal}{Journal of Optics}
  \textbf{\bibinfo{volume}{19}}, \bibinfo{pages}{044002}
  (\bibinfo{year}{2017}).

\end{thebibliography}
\end{document}